\documentclass[conference]{IEEEtran}
\ifCLASSINFOpdf
  \usepackage[pdftex]{graphicx}
\else
\fi
\usepackage{amsmath,amsfonts}
\usepackage{cite}
\usepackage{listings}
\usepackage{flushend}

\renewcommand{\matrix}[1]{{\bf #1}}

\begin{document}
\title{
Enabling Massive Deep Neural Networks \\
with the GraphBLAS}

\author{\IEEEauthorblockN{
  Jeremy Kepner$^1$,
  Manoj Kumar$^2$,
  Jos\'e Moreira$^2$, \\
  Pratap Pattnaik$^2$,
  Mauricio Serrano$^2$,
  Henry Tufo$^2$
}
\vspace{1ex}
\IEEEauthorblockA{
	\parbox{2.5in}{$^1$Massachusetts Institute of Technology \\ {\tt kepner@ll.mit.edu}}
	\parbox{4.0in}{$^2$IBM Thomas J. Watson Research Center \\ {\tt \{manoj1,jmoreira,pratap,mserrano,hmtufo\}@us.ibm.com}}
%%  $^3$MIT Mathematics Department,\\
%  $^g$Georgia Institute of Technology,
%  $^l$Lawrence Berkeley National Laboratory,\\
%  $^c$Carnegie Mellon University,
%  $^w$University of Washington,
%  $^i$IBM,
%  $^k$Karlsruhe Institute of Technology
%  $^d$
}
%\and
%\IEEEauthorblockN{Homer Simpson}
%\IEEEauthorblockA{Twentieth Century Fox\\
%\and
%\IEEEauthorblockN{James Kirk\\ and Montgomery Scott}
%\IEEEauthorblockA{Starfleet Academy\\
%San Francisco, California 96678--2391\\
%Fax: (888) 555--1212}
}


%Telephone: (800) 555--1212, Fax: (888) 555--1212}


\pagestyle{plain}

\maketitle

\begin{abstract}
Deep Neural Networks (DNNs) have emerged as a core tool for machine learning.
The computations performed during DNN training and inference are dominated by operations on
the weight matrices describing the DNN.  As DNNs incorporate more stages and more
nodes per stage, these weight matrices may be required to be sparse because of memory limitations.
% and because each internal node becomes more specialized, incorporating inputs from fewer nodes in the layer below.
The GraphBLAS.org math library standard was
developed to provide high performance manipulation of sparse weight matrices and input/output vectors.
For sufficiently sparse matrices, a sparse matrix library requires significantly less memory than the corresponding dense matrix implementation.
%and efficient methods to perform the underlying computations.
This paper provides a brief description of the mathematics underlying the GraphBLAS.  In addition,
the equations of a typical DNN are rewritten in a form designed
to use the GraphBLAS.   An implementation of the DNN is given
using a preliminary GraphBLAS C library.  The  performance of the
GraphBLAS implementation is measured relative to a standard dense linear algebra
library implementation. For various sizes of
DNN weight matrices, it is shown that the GraphBLAS sparse implementation
outperforms a BLAS dense implementation as the weight matrix becomes sparser.
%In this paper, we demonstrate that it can be used effectively to express computations underlying DNN training and inferencing.  Sparse matrix methods require significantly less memory than dense matrices.
%, unlike their dense matrix counterparts like BLAS, incur an overhead on per computation basis due to more complex data structures and algorithms needed to avoid computations with zero values.  Therefore, while they are less efficient than BLAS when computing with dense matrices, they
%outperform BLAS when computing with sparse matrices.  In this paper,
%we present experimental evidence of this tradeoff for DNN inferencing
%and the forward propagation step in DNN training.  For various sizes of
%DNN intra-stage weight matrices, we show that GraphBLAS implementation
%outperforms BLAS implementation as the weight matrix becomes sparse.
\end{abstract}


% \fi
\IEEEpeerreviewmaketitle

\thispagestyle{plain}

\section{Introduction}
\let\thefootnote\relax\footnotetext{This material is based in part upon
work supported by the NSF under grant number DMS-1312831.  Any opinions,
findings, and conclusions or recommendations expressed in this material
are those of the authors and do not necessarily reflect the views of
the National Science Foundation.}
Machine learning describes the broad area of analysis and classification
of data to create models for making predictions. Machine learning
has been the foundation of artificial intelligence since its inception
\cite{ware1955introduction,clark1955generalization,selfridge1955pattern,dinneen1955programming,newell1955chess,mccarthy2006proposal,minsky1960learning,minsky1961steps}.
Early machine learning applications included speech
recognition \cite{selfridge1955pattern}, computer vision
\cite{dinneen1955programming}, and even board games
\cite{newell1955chess,samuel1959some}.

\begin{figure}[htb]
  	\centering
    	\includegraphics[width=3in]{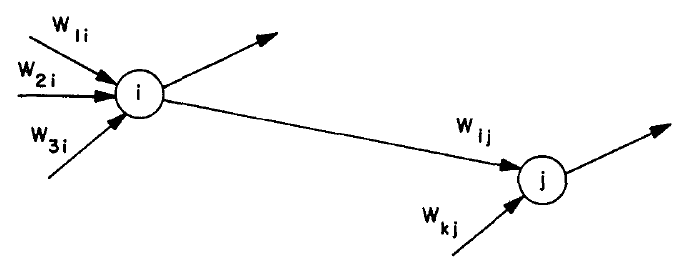}
      	\caption{Typical network elements $i$ and $j$ showing connection weights $w$ (reproduced from  \cite{clark1955generalization})}
      	\label{fig:clark1955fig1}
\end{figure}

\begin{figure}[htb]
  	\centering
    	\includegraphics[width=3in]{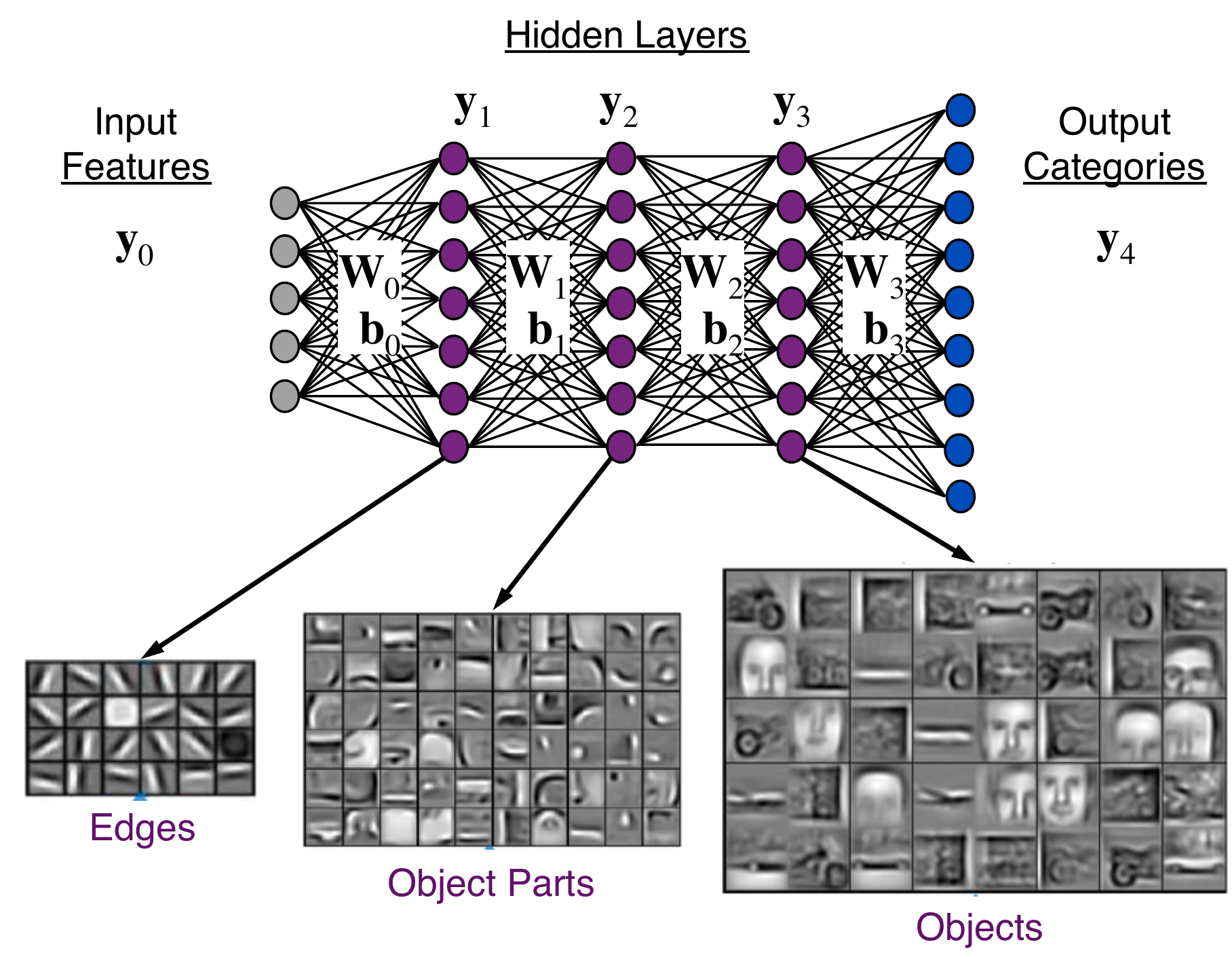}
	\caption{Four layer ($L=4$) deep neural network architecture
	for categorizing images.  The input
	features  ${\bf y}_0$ of an image are passed through a series
	of network layers ${\bf W}_{k=0,1,2,3}$, with bias terms
	${\bf b}_{k=0,1,2,3}$, that produce scores for categories
	${\bf y}_{L=4}$.  (Figure adapted from \cite{lee2009convolutional})}
      	\label{fig:DNNarchitecture}
\end{figure}

Using biological neuron inspired networks to implement machine learning
was the topic of the first paper presented at the first machine learning
conference in 1955 \cite{ware1955introduction,clark1955generalization}
(see Figure~\ref{fig:clark1955fig1}).  At this time, it was
recognized that direct computational training of neural networks was
computationally unfeasible \cite{minsky1960learning}.  The subsequent
many-fold improvement in neural network computation and theory has made it
possible to train neural networks that are capable of better-than-human
performance in a variety of important artificial intelligence problems
\cite{lippmann1987introduction,reynolds2000speaker,krizhevsky2012imagenet,lecun2015deep}.
Specifically, the availability of large corpora of validated data
sets \cite{campbell1995testing,lecun1998mnist,deng2009imagenet}
and the increases in computation spurred by games
\cite{campbell2002deep,mcgraw2007benchmarking,kerr2008gpu,epstein2012making},
have allowed the effective training of large deep neural networks
(DNNs) with 100,000s of input features, $N$, and hundreds of layers,
$L$, that are capable of choosing from among 100,000s categories, $M$
(see Figure~\ref{fig:DNNarchitecture}).

The impressive performance of large DNNs encourages the training and
testing of even larger networks.  However, increasing $N$, $L$, $M$
each by a factor 10 results in a 1000 fold increase in the memory
required for a DNN.  Because of these memory constraints, trade-offs
are currently being made in terms of precision and accuracy to save
storage and computation \cite{lavin2016fast,jouppi2017datacenter}.
Thus, there is significant interest in exploring the effectiveness of
sparse DNN representations where many of the weight values are zero.
As a comparison, the human brain has approximately 86 billion neurons and
150 trillion synapses~\cite{CNE:CNE21974}.  Its graph representation would
have approximately 2,000 edges per node, or a sparsity of $2 \times 10^3 /
86 \times 10^9 = 0.000002\%$.

If a large fraction of the DNN weights can be set to
zero, storage and computation costs can be reduced proportionately.
\cite{shi2017speeding,iandola2016squeezenet}.  The interest in
sparse DNNs is not limited to their computational advantages.
There has also been extensive theoretical work exploring the
potential neuromorphic and algorithmic benefits of sparsity
\cite{lee2008sparse,boureau2008sparse,glorot2011deep,yu2012exploiting}.
Experts in DNN believe that sparsification of weight matrices will lead
to better results.

As reported in the literature~\cite{DBLP:journals/corr/HanMD15}, sparsification of the weight
matrices is achieved by first training the neural network with a full
matrix, then removing those weights that are small, and finally retraining
the pruned network. The resulting sparse weight matrix can be
used during inference computation with the neural network.
Therefore, sparse matrix solutions can be used both during the 
final stages of training as well as during inference.

Computation over sparse data structures has been a mainstay
of the graph analysis community for many years.   Graphs are
among the most important abstract data structures in computer
science, and the algorithms that operate on them are critical to
applications in bioinformatics, computer networks, and social media
\cite{HendricksonKolda2000,Ediger2010,Ediger2011,Riedy2012,RiedyBader2013}.
Graphs have been shown to be powerful tools for modeling
complex problems because of their simplicity and generality
\cite{Bergamini2015,BergaminiMeyerhenke2016}.  For this reason, the field
of graph algorithms has become one of the pillars of theoretical computer
science, performing research in such diverse areas as combinatorial
optimization, complexity theory, and topology.  Graph algorithms have
been adapted and implemented by the military, commercial industry, and
researchers in academia, and have become essential in controlling the
power grid, telephone systems, and, of course, computer networks.

The connection between graphs and DNNs lies in the standard
matrix representation of graphs.  The duality between the
canonical representation of graphs as abstract collections of
vertices and edges and a matrix representation has been a part
of graph theory since its inception \cite{Konig1931, Konig1936}.
Matrix algebra has been recognized as a useful tool in graph
theory for nearly as long (see \cite{Harary1969} and references
therein, in particular \cite{Sabadusi1960,Weischel1962,McAndrew1963,
TehYap1964,McAndrew1965,HararyTauth1966,Brualdi1967}).  Likewise,
graph-based approaches have also been very useful in  matrix calculations
\cite{Parter1961,Fiedler1973,Gilbert1994}.  The modern description
of the duality between graph algorithms and matrix mathematics (or
sparse linear algebra) has been extensively covered in the literature
and is summarized in the cited text \cite{KepnerGilbert2011}.
This text has further spawned the development of the GraphBLAS
math library standard (GraphBLAS.org)\cite{Mattson2013}
that has been developed in a series of proceedings
\cite{Kepner2013gb,Mattson2014a,Mattson2014b,Mattson2015,Buluc2015,Mattson2016,BulucMcMillan2016,BulucMattson2017}
and implementations
\cite{BulucGilbert2011,Kepner2012-ch1,Ekanadham2014,Hutchison2015,Anderson2016,Wang2016,Zhang2016}.

In theory, the GraphBLAS may represent an ideal interface for
enabling massive DNNs on both conventional and custom hardware
\cite{Song2010,Song2013,song2016novel}.  Almost all computations
during DNN inferencing, and a significant fraction during training, are
encompassed in forward propagation of inferred features at each stage of
the network.  Intra-stage forward propagation consists of multiplication
of a weight matrix with a batch of input/output vectors, basically a
multiplication of two matrices, followed by an element wise application
of a non-linear function to the resulting matrix.  In this paper we show
that for sparse weight matrices these computations are performed much more
efficiently by GraphBLAS implementations than by BLAS implementations.

The rest of this paper explores this proposition as follows.  First, a
brief description of the mathematics underlying the GraphBLAS is provided.
Second, the mathematics of a common DNN are presented in a form designed
to utilize the  GraphBLAS.  Third, an implementation of the DNN is given
using a preliminary GraphBLAS C library.  Fourth, the  performance of
the GraphBLAS implementation is measured relative to standard linear algebra
library implementation.  Finally, this paper concludes with a summary
and recommendations on future work.

\section{GraphBLAS Mathematics}
  This section summarizes the GraphBLAS matrix mathematics relevant to the GraphBLAS DNN implementation.  For a more complete description of the mathematics in the GraphBLAS see \cite{KepnerGilbert2011,Kepner2016graphblas,kepner2017mathematics}. The foundational mathematical construct of matrix-based graph analysis is the adjacency matrix.  From this construct, a more general definition of a matrix can be constructed.  How such a matrix can be manipulated depends on the types of values in the matrix and the operations allowed on those values.  Furthermore, the mathematical properties of the matrix values determine the mathematical properties of the whole matrix.  Perhaps the most important aspect of the GraphBLAS mathematics is that it allows graph operations to be treated as \emph{Linear Systems}.  By exploiting the properties of linear systems, a GraphBLAS library can optimally order and even eliminate steps in a computation.
  
\subsection{Adjacency Matrix}
\begin{figure}[htb]
  \centering
    \includegraphics[width=3in]{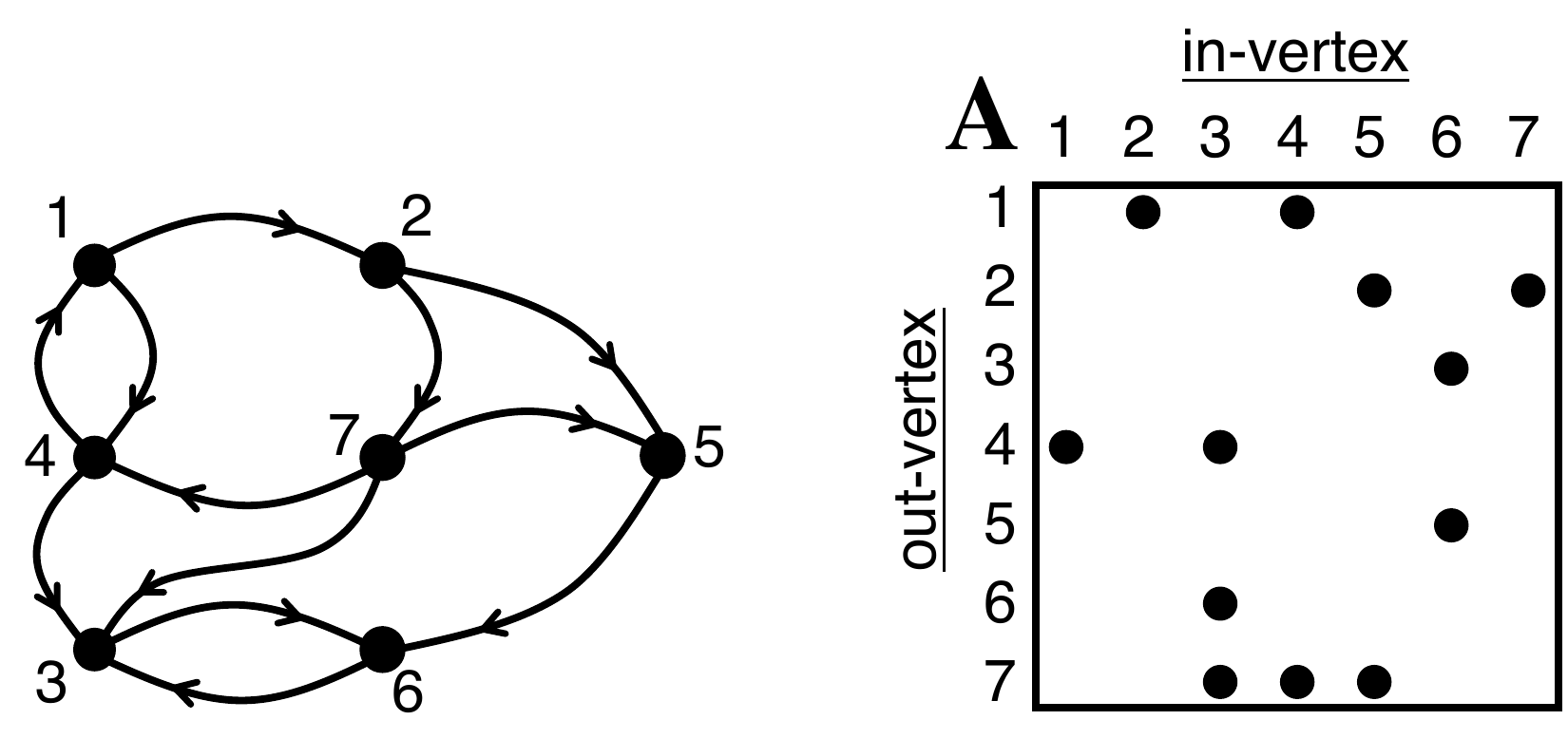}
      \caption{(left) Seven-vertex graph with 12 edges.  Each vertex is labeled with an integer.  (right)  $7 \times 7$ adjacency matrix $\mathbf{A}$ representation of the graph. $\mathbf{A}$ has 12 nonzero entries corresponding to the edges in the graph.}
      \label{fig:AdjacencyMatrix}
\end{figure}

Given an adjacency matrix $\mathbf{A}$, if
$$
  \mathbf{A}(i,j) = 1
$$
then there exists an edge going from vertex $i$ to vertex $j$ (see Figure~\ref{fig:AdjacencyMatrix}).  Likewise, if
$$
  \mathbf{A}(i,j) = 0
$$
then there is no edge from $i$ to $j$.  Adjacency matrices can have direction, which means that $\mathbf{A}(i,j)$ may not be the same as $\mathbf{A}(j,i)$.  Adjacency matrices can also have edge weights.  If
$$
  \mathbf{A}(i,j) = a \neq 0
$$
then the edge going from $i$ to $j$ is said to have weight $a$. Adjacency matrices provide a simple way to represent the connections between vertices in a graph.  Adjacency matrices are often square, and both the out-vertices (rows) and the in-vertices (columns) are the same set of vertices.  Adjacency matrices can be rectangular, in which case the out-vertices (rows) and the in-vertices (columns) are different sets of vertices.  Such graphs are often called bipartite graphs.  In summary, adjacency matrices can represent a wide range of graphs, which include any graph with any set of the following properties: directed, weighted, and/or bipartite.

\subsection{Matrix Values}
  A typical matrix has $m$ rows and $n$ columns of real numbers.  Such a matrix can be denoted as
$$
  \mathbf{A}: \mathbb{R}^{m \times n}
$$
The row and and column indexes of the matrix $\mathbf{A}$ are
$$
  i \in I = \{1,\ldots,m\}
$$
and
$$
  j \in J = \{1,\ldots,n\}
$$
so that any particular value $\mathbf{A}$ can be denoted as $\mathbf{A}(i,j)$.   The row and column indices of matrices are natural numbers $I,J : \mathbb{N}$.
A matrix of integers is denoted
$
  \mathbf{A}: \mathbb{Z}^{m \times n}
$, and
  a matrix of natural numbers is denoted
$
  \mathbf{A}: \mathbb{N}^{m \times n}
$.
Using the above concepts, a matrix is defined as the following two-dimensional (2D) mapping
$$
  \mathbf{A} : I \times J \rightarrow \mathbb{S}
$$
where the indices $I, J : \mathbb{Z}$ are finite sets of integers with $m$ and $n$ elements, respectively, and
$$
  \mathbb{S} \in \{\mathbb{R},\mathbb{Z},\mathbb{N}, \ldots \}
$$
is a set of scalars.  Without loss of generality, matrices can be denoted
$$
  \mathbf{A}: \mathbb{S}^{m \times n}
$$
A \emph{vector} is a matrix in which either $m=1$ or $n=1$. A column vector is denoted
$
 \mathbf{v} : \mathbb{S}^{m \times 1}
$
or simply
$
 \mathbf{v} : \mathbb{S}^{m}
$.
A row vector can be denoted
$
  \mathbf{v} : \mathbb{S}^{1 \times n}
$
or simply 
$
  \mathbf{v} : \mathbb{S}^{n}
$.
A scalar is a single element of a set
$
  s \in \mathbb{S}
$
and has no matrix dimensions.

\subsection{Scalar Operations}
Matrix operations are built on top of scalar operations that can be used for combining and scaling graph edge weights.  The primary scalar operations are standard arithmetic addition, such as
$$
  1 + 1 = 2
$$
and arithmetic multiplication, such as
$$
  2 \times 2 = 4
$$
These scalar operations of addition and multiplication can be defined to be a wide variety of functions.  To prevent confusion with standard arithmetic addition and arithmetic multiplication, $\oplus$ will be used to denote scalar addition and $\otimes$ will be used to denote scalar multiplication.  In this notation, standard arithmetic addition and arithmetic multiplication of real numbers
$$
  a, b, c \in \mathbb{R}
$$
where
$$
  \oplus \equiv + ~~~~~ \text{and} ~~~~~ \otimes \equiv \times
$$
results in
$$
   c = a \oplus b  ~~~~~~~~~ \Rightarrow ~~~~~~~~~ c = a + b
$$
and
$$
   c = a \otimes b  ~~~~~~~~~ \Rightarrow ~~~~~~~~~ c = a \times b
$$
Generalizing $\oplus$ and $\otimes$ to a variety of operations enables a wide range of algorithms on scalars of all different types (not just real or complex numbers).

Certain $\oplus$ and $\otimes$ combinations  over certain sets of scalars are particularly useful, and referred to as \emph{semirings}, because they preserve essential mathematical properties, such as
additive commutativity
$$
  a \oplus b = b \oplus a
$$
%multiplicative commutativity \comment{do semirings really need multiplicative commutativity?}
%$$
%  a \otimes b = b \otimes a
%$$
additive associativity
$$
  (a \oplus b) \oplus c = a \oplus (b \oplus c)
$$
multiplicative associativity
$$
  (a \otimes b) \otimes c = a \otimes (b \otimes c)
$$
and the distributivity of multiplication over addition
$$
  a \otimes (b \oplus c)  = (a \otimes b) \oplus (a \otimes c)
$$

The properties of commutativity, associativity, and distributivity are \emph{extremely} useful properties for building graph applications because they allow the builder to swap operations without changing the result.  Example combinations of $\oplus$ and $\otimes$ that preserve scalar commutativity, associativity, and distributivity include (but are not limited to) standard arithmetic
$$
  \oplus \equiv + ~~~~~~~~~ \otimes \equiv \times ~~~~~~~~~ a, b, c \in \mathbb{R}
$$
max-plus algebras
$$
  \oplus \equiv \max ~~~~~~~~~ \otimes \equiv + ~~~~~~~~~ a, b, c \in \{-\infty \cup \mathbb{R}\}
$$
max-min algebras
$$
  \oplus \equiv \max ~~~~~~~~~ \otimes \equiv \min ~~~~~~~~~ a, b, c \in \{\text{-}\infty \cup \mathbb{R}_{\leq 0} \}
$$
finite (Galois) fields such as GF(2)
$$
  \oplus \equiv {\rm xor} ~~~~~~~~~ \otimes \equiv {\rm and} ~~~~~~~~~ a, b, c \in \{0,1\}
$$
and power set algebras
$$
  \oplus \equiv \cup ~~~~~~~~~ \otimes \equiv \cap ~~~~~~~~~ a, b, c \subset \mathcal{P}(\mathbb{Z})
$$
%Other functions that do not preserve the above properties can also be defined for $\oplus$ and $\otimes$.  For example, it is often useful for $\oplus$ or $\otimes$ to pull in other data, such as vertex indices of a graph.

\subsection{Matrix Properties}
\label{MatrixProperties}
Associativity, distributivity, and commutativity are very powerful properties that enable the construction of composable graph algorithms (i.e., operations can be reordered with the knowledge that the answers will remain unchanged).  Composability makes it easy to build a wide range of graph algorithms with just a few functions.  Given matrices
$$
  \mathbf{A}, \mathbf{B}, \mathbf{C} \in \mathbb{S}^{m \times n}
$$
let their elements be specified by
$$
  a = \mathbf{A}(i,j) ~~~~~ 
  b = \mathbf{B}(i,j) ~~~~~
  c = \mathbf{C}(i,j)
$$
Commutativity, associativity, and distributivity of scalar operations translates into similar properties on matrix operations in the following manner.

Element-wise additive commutativity of matrices 
  $$
      a \oplus b = b \oplus a  ~~~~~ \Rightarrow ~~~~~
      \mathbf{A} \oplus \mathbf{B} = \mathbf{B} \oplus \mathbf{A}
  $$
  where matrix element-wise addition is given by
  $$
      \mathbf{C}(i,j) = \mathbf{A}(i,j) \oplus \mathbf{B}(i,j)
  $$
Element-wise multiplicative commutativity of matrices
  $$
      a \otimes b = b \otimes a  ~~~~~ \Rightarrow ~~~~~
      \mathbf{A} \otimes \mathbf{B} = \mathbf{B} \otimes \mathbf{A}
  $$
    where matrix element-wise (Hadamard) multiplication is given by
  $$
       \mathbf{C}(i,j) = \mathbf{A}(i,j) \otimes \mathbf{B}(i,j)
  $$
Element-wise additive associativity of matrices
  \vspace{3pt}\\
  $~~~~~~~~~~~~~~~~~
      (a \oplus b) \oplus c = a \oplus (b \oplus c) 
  $
  \\
  $~~~~~ \Rightarrow ~~~~~
      (\mathbf{A} \oplus \mathbf{B}) \oplus \mathbf{C} = \mathbf{A} \oplus (\mathbf{B} \oplus \mathbf{C})
  $
  \vspace{3pt}\\
%  $$
%      (a \oplus b) \oplus c = a \oplus (b \oplus c)   \Rightarrow
%      (\mathbf{A} \oplus \mathbf{B}) \oplus \mathbf{C} = \mathbf{A} \oplus (\mathbf{B} \oplus \mathbf{C})
%  $$
Element-wise multiplicative associativity of matrices
  \vspace{3pt}\\
  $~~~~~~~~~~~~~~~~~
      (a \otimes b) \otimes c = a \otimes (b \otimes c) 
  $
  \\
  $~~~~~ \Rightarrow ~~~~~
      (\mathbf{A} \otimes \mathbf{B}) \otimes \mathbf{C} = \mathbf{A} \otimes (\mathbf{B} \otimes \mathbf{C})
  $
  \vspace{3pt}\\
%  $$
%      (a \otimes b) \otimes c = a \otimes (b \otimes c)   \Rightarrow
%      (\mathbf{A} \otimes \mathbf{B}) \otimes \mathbf{C} = \mathbf{A} \otimes (\mathbf{B} \otimes \mathbf{C})
%  $$
Element-wise distributivity of matrices
  \vspace{3pt}\\
  $~~~~~~~~~~~~~~~~~
      a \otimes (b \oplus c) = (a \otimes b) \oplus (a \otimes c) 
  $
  \\
  $~~~~~ \Rightarrow ~~~~~
      \mathbf{A} \otimes (\mathbf{B} \oplus \mathbf{C}) = (\mathbf{A} \otimes \mathbf{B}) \oplus (\mathbf{A} \otimes \mathbf{C})
  $
  \vspace{3pt}\\
%  $$
%      a \otimes (b \oplus c) = (a \otimes b) \oplus (a \otimes c)   \Rightarrow
%      \mathbf{A} \otimes (\mathbf{B} \oplus \mathbf{C}) = (\mathbf{A} \otimes \mathbf{B}) \oplus (\mathbf{A} \otimes \mathbf{C})
%  $$
Matrix multiply distributivity
  \vspace{3pt}\\
  $~~~~~~~~~~~~~~~~~
      a \otimes (b \oplus c) = (a \otimes b) \oplus (a \otimes c) 
  $
  \\
  $~~~~~ \Rightarrow ~~~~~~~~
      \mathbf{A} (\mathbf{B} \oplus \mathbf{C}) = (\mathbf{A} \mathbf{B}) \oplus (\mathbf{A} \mathbf{C})
  $
  \vspace{3pt}\\
%  $$
%      a \otimes (b \oplus c) = (a \otimes b) \oplus (a \otimes c)  \Rightarrow
%      \mathbf{A} (\mathbf{B} \oplus \mathbf{C}) = (\mathbf{A} \mathbf{B}) \oplus (\mathbf{A} \mathbf{C})
%  $$
 where matrix multiply
 $$
   \mathbf{C} = \mathbf{A} {\oplus}.{\otimes} \mathbf{B} = \mathbf{A} \mathbf{B}
$$
is given by
  $$
   {\bf C}(i,j) = \bigoplus_{k=1}^l {\bf A}(i,k) \otimes {\bf B}(k,j)
  $$
for matrices with dimensions
$$
  {\bf A} : \mathbb{S}^{m \times l} ~~~~~
  {\bf B} : \mathbb{S}^{l \times m} ~~~~~
  {\bf C} : \mathbb{S}^{m \times n}
$$
Matrix multiply associativity of matrices
  \vspace{3pt}\\
  $~~~~~~~~~~~~~~~~~
      a \otimes (b \oplus c) = (a \otimes b) \oplus (a \otimes c) 
  $
  \\
  $~~~~~ \Rightarrow ~~~~~~~~~~~
      (\mathbf{A} \mathbf{B}) \mathbf{C} = \mathbf{A} (\mathbf{B} \mathbf{C})
  $
  \vspace{3pt}\\
%  $$
%      a \otimes (b \oplus c) = (a \otimes b) \oplus (a \otimes c)  \Rightarrow 
%      (\mathbf{A} \mathbf{B}) \mathbf{C} = \mathbf{A} (\mathbf{B} \mathbf{C})
%  $$
%Matrix multiply commutativity can be achieved when combined with the transpose operation
%$$
%  (\mathbf{A} \mathbf{B})^{\sf T} = \mathbf{B}^{\sf T} \mathbf{A}^{\sf T}
%$$
%where the transpose of a matrix is given by
%$$
%  \mathbf{A}^{\sf T}(j,i) = \mathbf{A}(i,j)
%$$

\subsection{0-Element: No Graph Edge}
Sparse matrices play an important role in graphs.  Many implementations of sparse matrices reduce storage by not storing the 0-valued elements in the matrix.  In adjacency matrices, the 0 element is equivalent to no edge from the vertex that is represented by the row to the vertex that is represented by the column. In incidence matrices, the 0 element is equivalent to the edge represented by the row not including the vertex that is represented by the column.  In most cases, the 0 element is standard arithmetic 0, but in other cases it can be a different value.  Nonstandard 0 values can be helpful when combined with different $\oplus$ and $\otimes$ operations.  For example, in different contexts 0 might be $+\infty$, -$\infty$, or $\emptyset$ (empty set).
For any value of 0, if the 0 element has certain properties with respect to scalar $\oplus$ and $\otimes$, then the sparsity of matrix operations can be managed efficiently.  These properties are the additive identity
$$
     a \oplus 0 = a
$$
and the multiplicative annihilator
$$
     a \otimes 0 = 0
$$

Example combinations of $\oplus$ and $\otimes$ that exhibit the additive identity and multiplicative annihilator include standard arithmetic  (${+}.{\times}$) on real numbers $\mathbb{R}$, max-plus algebra (${\max}.{+}$) on real numbers with a defined minimal element $\{\text{-}\infty \cup \mathbb{R}\}$, and min-max algebra (${\min}.{\max}$) using non-negative real numbers with a maximal element $\{\mathbb{R}_{\geq 0} \cup \infty\}$.
%\begin{itemize}
%\item standard arithmetic  (${+}.{\times}$) on real numbers $\mathbb{R}$
%\item max-plus algebra (${\max}.{+}$) on real numbers with a defined minimal element $\{\text{-}\infty \cup \mathbb{R}\}$
%\item min-plus algebra (${\min}.{+}$) using real numbers with a defined maximal element $\{\mathbb{R} \cup \infty\}$
%\item max-min algebra (${\max}.{\min}$) using non-negative real numbers $[0,\infty)$
%\item min-max algebra (${\min}.{\max}$)] using non-positive real numbers $(\text{-}\infty,0]$
%\item max-min algebra (${\max}.{\min}$) using non-positive real numbers with a minimal element $\{\text{-}\infty \cup \mathbb{R}_{\leq 0} \}$
%\item min-max algebra (${\min}.{\max}$) using non-negative real numbers with a maximal element $\{\mathbb{R}_{\geq 0} \cup \infty\}$
%\item Galois field (${{\rm xor}}.{{\rm and}}$) over a set of two numbers $\{0,1\}$
%\item power set (${\cup}.{\cap}$)] on any subset of integers $\mathbb{Z}$
%\end{itemize}
The above examples are a small selection of the operators and sets that are useful for building graph algorithms.  Many more are possible.  The ability to change the  scalar values and operators while preserving the overall behavior of the graph operations is one of the principal benefits of using matrices for graph algorithms.

\section{Deep Neural Network Mathematics}

  The primary mathematical operation performed by a DNN network is the inference, or forward propagation, step.  Inference is executed repeatedly during training to determine both the weight matrices ${\bf W}_k$ and the bias vectors ${\bf b}_k$ of the DNN.  The inference computation shown in Figure~\ref{fig:DNNarchitecture} is given by
$$
  {\bf y}_{k + 1} = h({\bf W}_k {\bf y}_k + {\bf b}_k)
$$
where $h()$ is a non-linear function applied to each element of the vector.  A commonly used function is the rectified linear unit (ReLU) given by
$$
   h({\bf y}) = \max({\bf y},0)
$$
which sets values less that 0 to 0 and leaves other values unchanged.  When training a DNN, it is common to compute multiple ${\bf y}_k$ vectors at once in a batch that can be denoted as the matrix ${\bf Y}_k$.  In matrix form, the inference step becomes
$$
  {\bf Y}_{k + 1} = h({\bf W}_k {\bf Y}_k + {\bf B}_k)
$$
where ${\bf B}_k$ is a replication of ${\bf b}_k$ along columns.

  If $h()$ were a linear function, then the above equation could be solved exactly and the computation could be greatly simplified.  However, current evidence suggests that the non-linearity of $h()$ is required for a DNN to be effective.  Interestingly, the inference computation can be rewritten as a linear function over two different semirings
$$
  {\bf y}_{k + 1} ={\bf W}_k {\bf y}_k \otimes {\bf b}_k \oplus 0
$$
or in matrix form
$$
  {\bf Y}_{k + 1} ={\bf W}_k {\bf Y}_k \otimes {\bf B}_k \oplus 0
$$
where the $\oplus = \max$ and $\otimes = +$.  Thus, ${\bf W}_k {\bf y}_k$ and ${\bf W}_k {\bf Y}_k$ are computed over the standard arithmetic ${+}.{\times}$ semiring 
$$
  S_1 = (\mathbb{R},+,\times,0,1)
$$
while the $\oplus$ and $\otimes$  operation are performed over the ${\max}.{+}$ semiring
$$
  S_2 = (\{ \text{-}\infty \cup \mathbb{R} \},\max,+,\text{-}\infty,0)
$$
Thus, the ReLU DNN can be written as a linear system that oscillates over two semirings $S_1$ and $S_2$.  $S_1$ is the most widely used of semirings and performs standard correlation between vectors.  $S_2$ is also a commonly used semiring for selecting optimal paths in graphs.  Thus, the inference step of a ReLU DNN can be viewed as combining correlations of inputs and to choose optimal paths  through the neural network.  Finally, and perhaps most importantly, the GraphBLAS is designed to support the above semiring calculations over sparse matrices.

\section{Deep Neural Network Implementation}

The GraphBLAS C implementation of the above deep neural networks
is extremely concise, as shown in Figure~\ref{Fig:GraphBLASCode}.
Function {\sf dnn} (line 4) computes a forward propagation step in
a ReLU DNN with $L$ layers.  We assume that all layers have the
same number of neurons, $m$, which is not a big loss of generality
when the matrices can be sparse. The forward step is computed for a
minibatch of size $n$.  ${\bf W}$ is an array of $m \times m$ weight matrices, where ${\bf W}[k]$ is the weight matrix for layer
$k$. Correspondingly, ${\bf B}$ is an array of $m \times  n$-element bias
vectors, where ${\bf B}[k]$ is the bias for layer $k$. Finally, ${\bf Y}$
is an array of $m \times n$ matrices, where ${\bf Y}[k]$ is
the output from layer $k-1$ and the input for layer $k$. (${\bf Y}[0]$
is the input and ${\bf Y }[L]$ the output for the entire network.)

\begin{figure*}
	{\scriptsize
\begin{lstlisting}[language=C,numbers=left]
#include <math.h>
#include <GraphBLAS.h>

GrB_Info dnn(GrB_Matrix *Y, GrB_Matrix *W, GrB_Matrix *B, GrB_Index L)
/*
 * L        - Number of layers
 * W[0:L-1] - Array of m x m weight matrices
 * B[0:L-1] - Array of m x n bias matrices
 * Y[0:L]   - Array of m x n layer-input/output matrices
 */
{
  GrB_Monoid FP32Add;								// Monoid <float,+,0.0>
  GrB_Monoid_new(&FP32Add,GrB_FP32,GrB_PLUS_FP32,0.0f);
  GrB_Semiring FP32AddMul;							// Semiring <float,+,*,0.0>
  GrB_Semiring_new(&FP32AddMul,FP32Add,GrB_TIMES_FP32);

  GrB_Monoid FP32Max;								// Monoid <float,max,-inf>
  GrB_Monoid_new(&FP32Max,GrB_FP32,GrB_MAX_FP32,-INFINITY);
  GrB_Semiring FP32MaxPlus;							// Semiring <float,max,+,-inf>
  GrB_Semiring_new(&FP32MaxPlus,FP32Max,GrB_PLUS_FP32);

  GrB_Index m, n;      
  GrB_Matrix_nrows(&m,Y[0]); GrB_Matrix_ncols(&n,Y[0]);
  GrB_Matrix Zero;								// Zero = 0.0
  GrB_Matrix_new(&Zero,GrB_FP32,m,n);   
  GrB_assign(Zero,GrB_NULL,GrB_NULL,0.0,GrB_ALL,m,GrB_ALL,n,GrB_NULL);

  for(int k=0; k<L; k++)
  {
    GrB_mxm(Y[k+1],GrB_NULL,GrB_NULL,FP32AddMul,W[k],Y[k],GrB_NULL);		// Y[k+1] = W[k]*Y[k]
    GrB_eWiseMult(Y[k+1],GrB_NULL,GrB_NULL,FP32MaxPlus,Y[k+1],B[k],GrB_NULL);	// Y[k+1] = W[k]*Y[k] + B[k]
    GrB_eWiseAdd (Y[k+1],GrB_NULL,GrB_NULL,FP32MaxPlus,Y[k+1],Zero,GrB_NULL);	// Y[k+1] = max(W[k]*Y[k] + B[k],0)
  }

  GrB_free(&Zero);
  GrB_free(&FP32Add);    GrB_free(&FP32Max);
  GrB_free(&FP32AddMul); GrB_free(&FP32MaxPlus);

  return GrB_SUCCESS;
}
\end{lstlisting}}
	\caption{GraphBLAS implementation of ReLU DNN using the C API.}
	\label{Fig:GraphBLASCode}
\end{figure*}

Lines 12--15 and 17--20 define the two semirings that we use: 
the arithmetic semiring ({\sf FP32AddMul}) and
the max-plus semiring ({\sf FP32MaxPlus}).
(In GraphBLAS, semirings are built from monoids and binary operators.)

Lines 22--23 extract
the parameters $m$ and $n$ from matrix ${\bf Y}[0]$, which are
the same for all layers.  Lines 24--26 create an $m \times n$
matrix of zeros that will be used in the computation of the ReLU function.
Lines 35--37 free the GraphBLAS objects created inside this function.

The main computation loop is in lines 28--33. Each iteration computes
the forward step in one layer $k$ of the network.
The matrix-multiply in line 30 computes the product of the
weight matrix ${\bf W}[k]$ by the input matrix ${\bf Y}[k]$.
That result is added to the bias matrix ${\bf B}[k]$ in line 31. ($+$ is the multiplicative operation of semiring {\sf FPR32MaxPlus}.)
Finally, line 32 performs the ReLU operation by selecting the maximum
between the previous result and the zero matrix. ($\max$ is the additive operation of semiring {\sf FP32MaxPlus}.)

\section{Experimental Evaluation}

% We report performance results from various experiments with the GraphBLAS
% implementation of the ReLU DNN previously discussed, and compare that
% performance with a traditional implementation of the same DNNs through
% dense linear algebra. We describe our experimental platform, the different
% middleware we use, and the various matrices used in our experiments.
In this section we compare the performance of GraphBLAS implementation of
the forward propagation calculations in ReLU DNN with the OpenBLAS 
implementation.  As expected, when the weight matrix is dense, 
OpenBLAS outperforms the GraphBLAS implementation.  The converse is true
for sparse weight matrices.

\subsection{Experimental Platform and Middleware}

All our experiments are performed on a IBM Power S824L server
configured with 24 POWER8 cores running at 3.325
GHz. Each core is capable of running from 1 (single-thread or ST) to 8
(SMT8) simultaneous threads of execution.  The server is configured
with 512~GiB of memory, accessible through a total of 16 memory
channels. Total bandwidth from processing cores to memory is over 400
GB/s, with two-thirds of that for reading from memory and one-third for
writing to memory.

The operating system installed is Ubuntu 16.04 distribution of Linux
for PowerPC Little Endian.  All code as compiled with Gnu 5.4
compilers ({\sf gcc}/{\sf g++}). The GraphBLAS API was implemented
as a compatibility layer on top of IBM's Graph Processing Interface
(GPI) library~\cite{Horn2017}. When parallel processing was used in GraphBLAS, it was
accomplished transparently through GPI, which in turns relies on OpenMP
for multithreaded processing. GPI transparently uses various storage formats and strategies for dividing the work among multiple threads. In our specific
case, the weight matrices were represented in compressed sparse row (CSR) format and they were distributed by rows.
For dense linear algebra, we use OpenBLAS
version 0.2.18.

\subsection{Experimental Matrices}

All weight matrices $\matrix{W}$ are $m \times m$ square matrices of
single-precision (32-bit) floating-point numbers. All layer-input/output
matrices $\matrix{Y}$ are \emph{tall} and \emph{skinny} $m \times 64$
matrices that represent a mini-batch of size 64.

We vary the size ($m$) as well as the \emph{sparsity} of the matrices.
For convenience of presentation, we define the \emph{inverse sparsity}
of a matrix as the total number of elements in the matrix divided by the
number of nonzero elements. In other words, the larger the inverse sparsity
of a matrix, the more sparse it is (the larger the fraction of zeros).
When performing the dense linear algebra version of the computation,
zeros are represented explicitly. When using GraphBLAS, sparse matrix
representations are used and only the nonzeros are stored.

For the weight matrices $\matrix{W}$ we vary the inverse sparsity all
the way from 1 (dense matrix) to 262144 (only $0.0004\%$ of elements
are nonzero). Although it would be possible to treat the input/output
matrices as sparse, in our work we only consider
dense $\matrix{Y}$ matrices.

Initially, dense weight matrices are generating by populating each entry
with a random number chosen from 
a $U[-1,3)$ distribution. Sparse
weight matrices are generated from these dense matrices by selecting
the location of nonzero entries using independent Bernoulli distributions
and taking the corresponding entry value (generated from the $U[-1,3)$
distribution).
Layer input matrices are generated using a $U[0,1)$
distribution for the entry values.

\subsection{Performance Measurements}

\begin{figure*}
	\begin{center}
	\begin{tabular}{c}
		\includegraphics[width=5.0in]{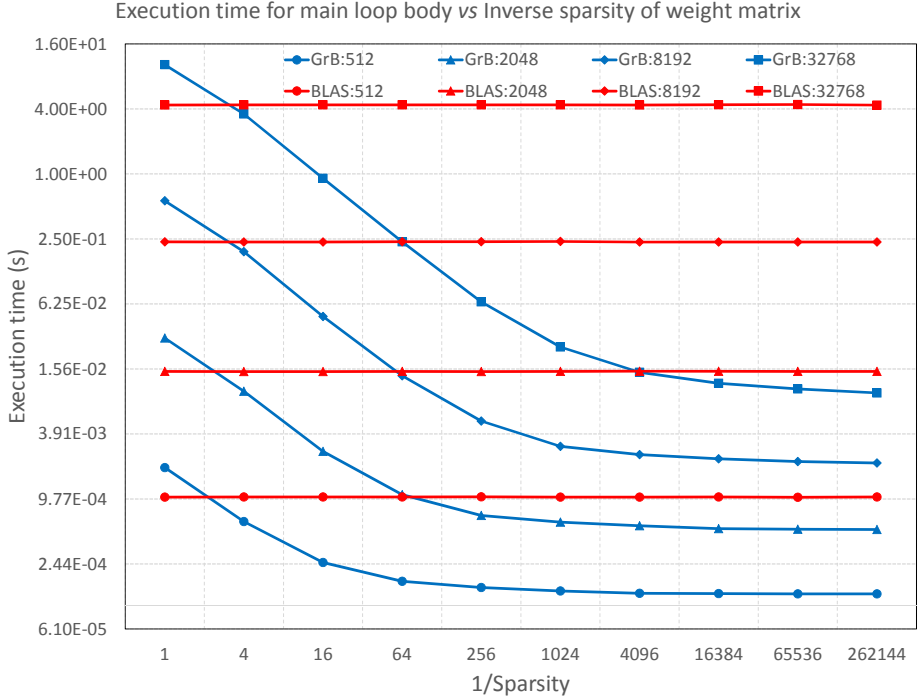} 
	\end{tabular}
	\end{center}
	\caption{Single-threaded (ST) results for GraphBLAS ({\sf GrB})
	and dense linear algebra ({\sf BLAS}) implementations of the ReLU
	DNN. Results for GraphBLAS are in blue, whereas results for BLAS
	are in red. Each problem size uses a different marker. (Compare
	blue and red curves with the same marker.)}
	\label{Fig:STresults}
\end{figure*}

Figure~\ref{Fig:STresults} shows the single-threaded results for both
GraphBLAS and dense linear algebra implementations of the ReLU DNN.
We plot results for four different values of the matrix size parameter
$n$ ($512$, $2048$, $8192$ and $32768$).  For each matrix size, we show
results for the two implementation versions ({\sf GrB} - GraphBLAS,
{\sf BLAS} - dense linear algebra).  The $x$-axis is the inverse sparsity
of $\matrix{W}$ and the $y$-axis is the average execution time of the
main loop iteration in Figure~\ref{Fig:GraphBLASCode}. Both axes use a
logarithmic scale.

\begin{figure*}
	\begin{center}
	\begin{tabular}{cc}
		\includegraphics[width=3.0in]{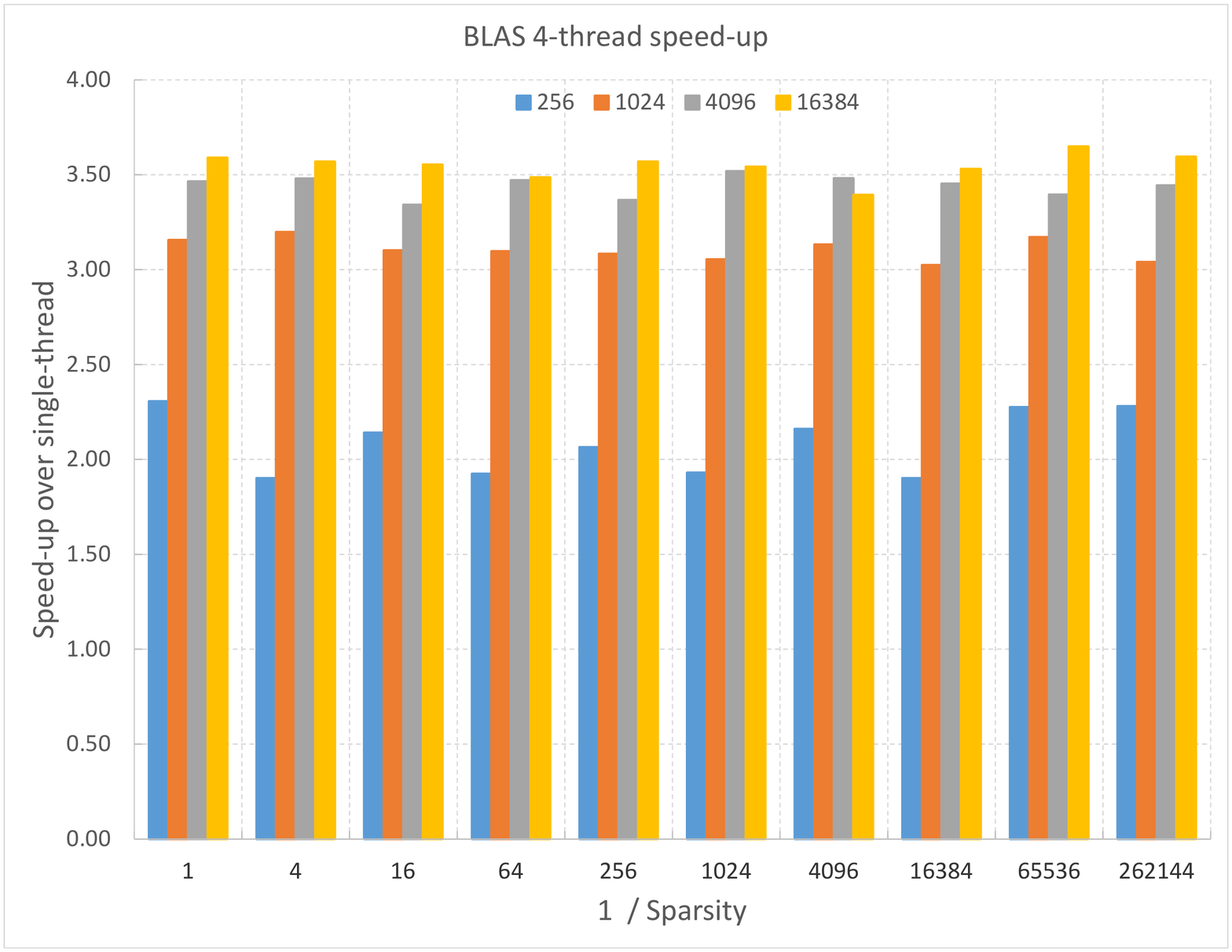} & \includegraphics[width=3.0in]{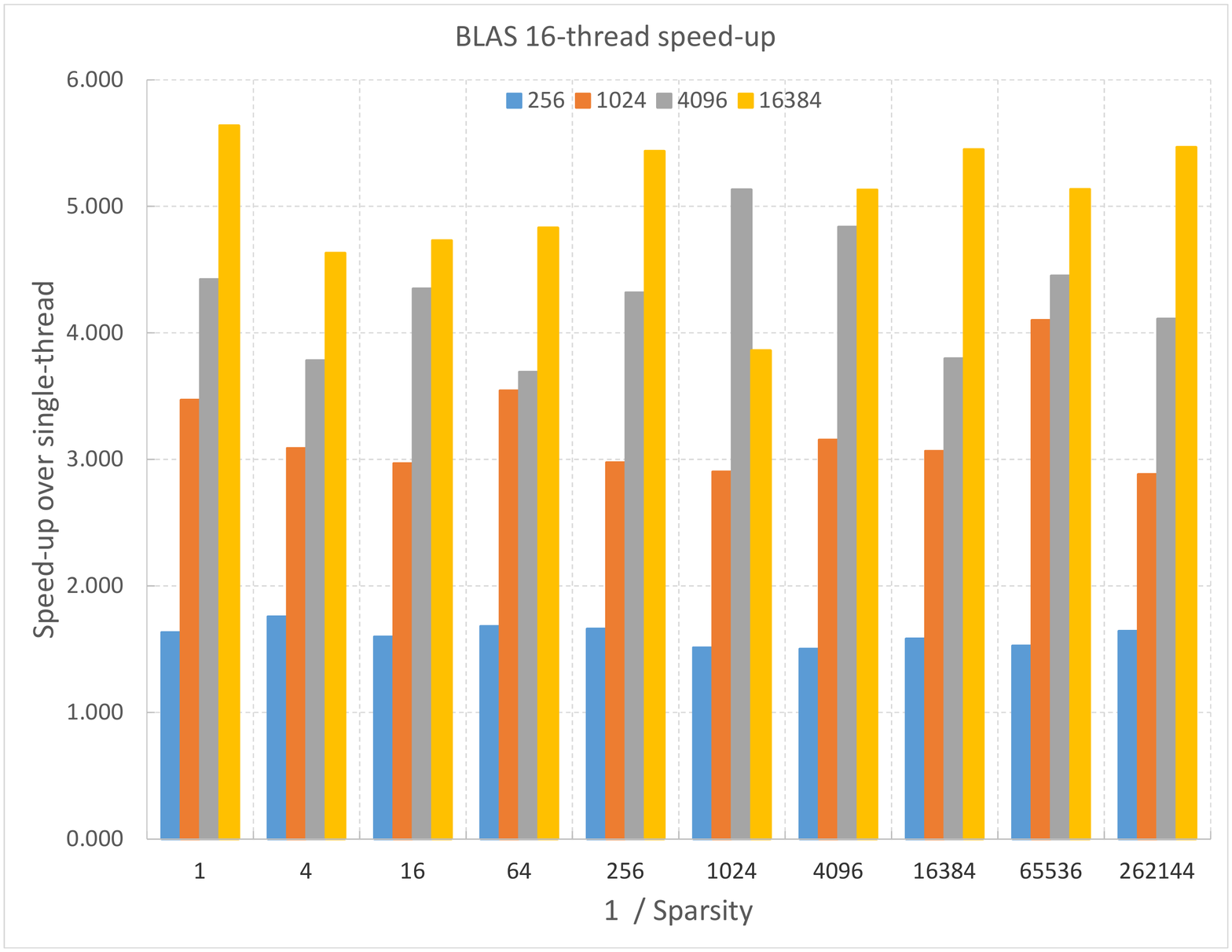} \\
		\includegraphics[width=3.0in]{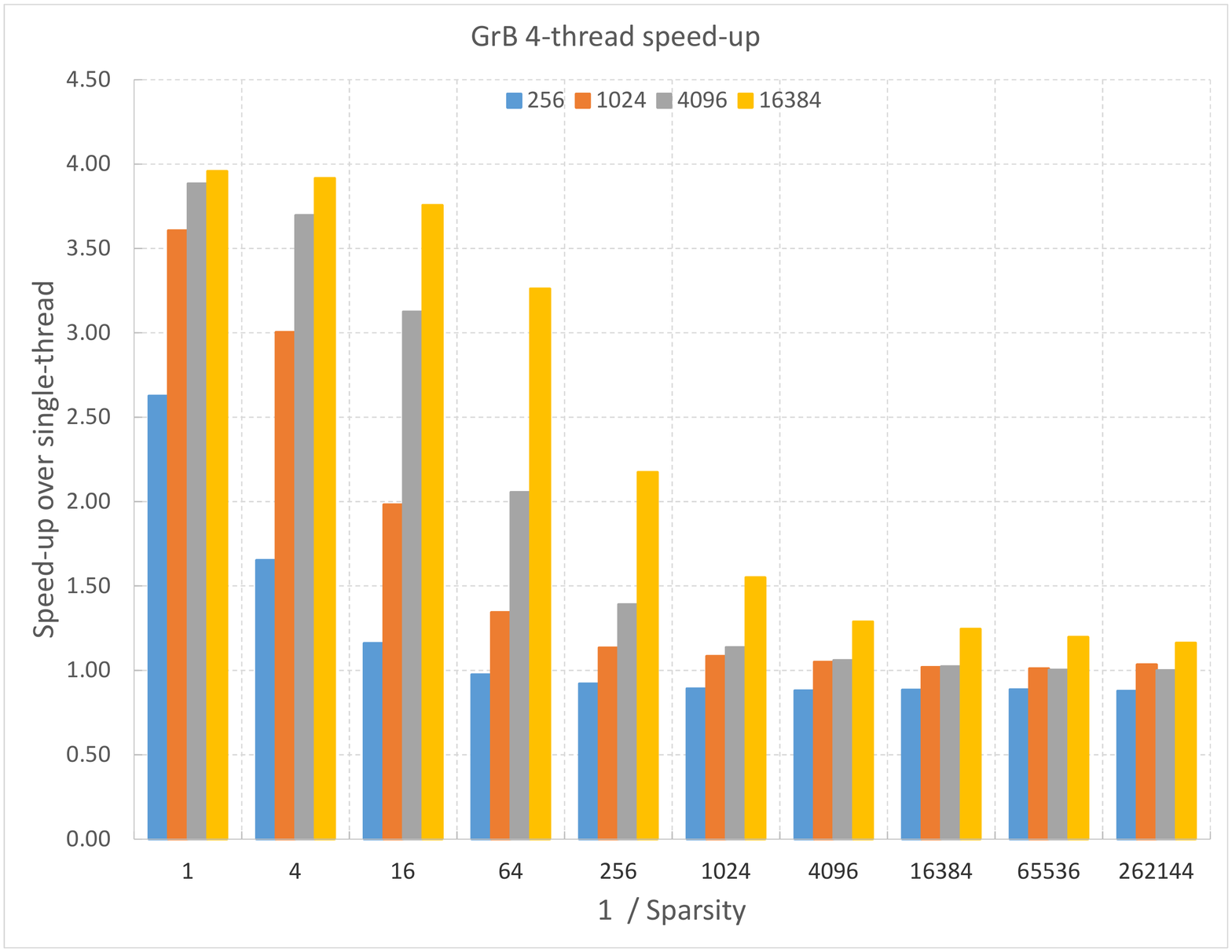} & \includegraphics[width=3.0in]{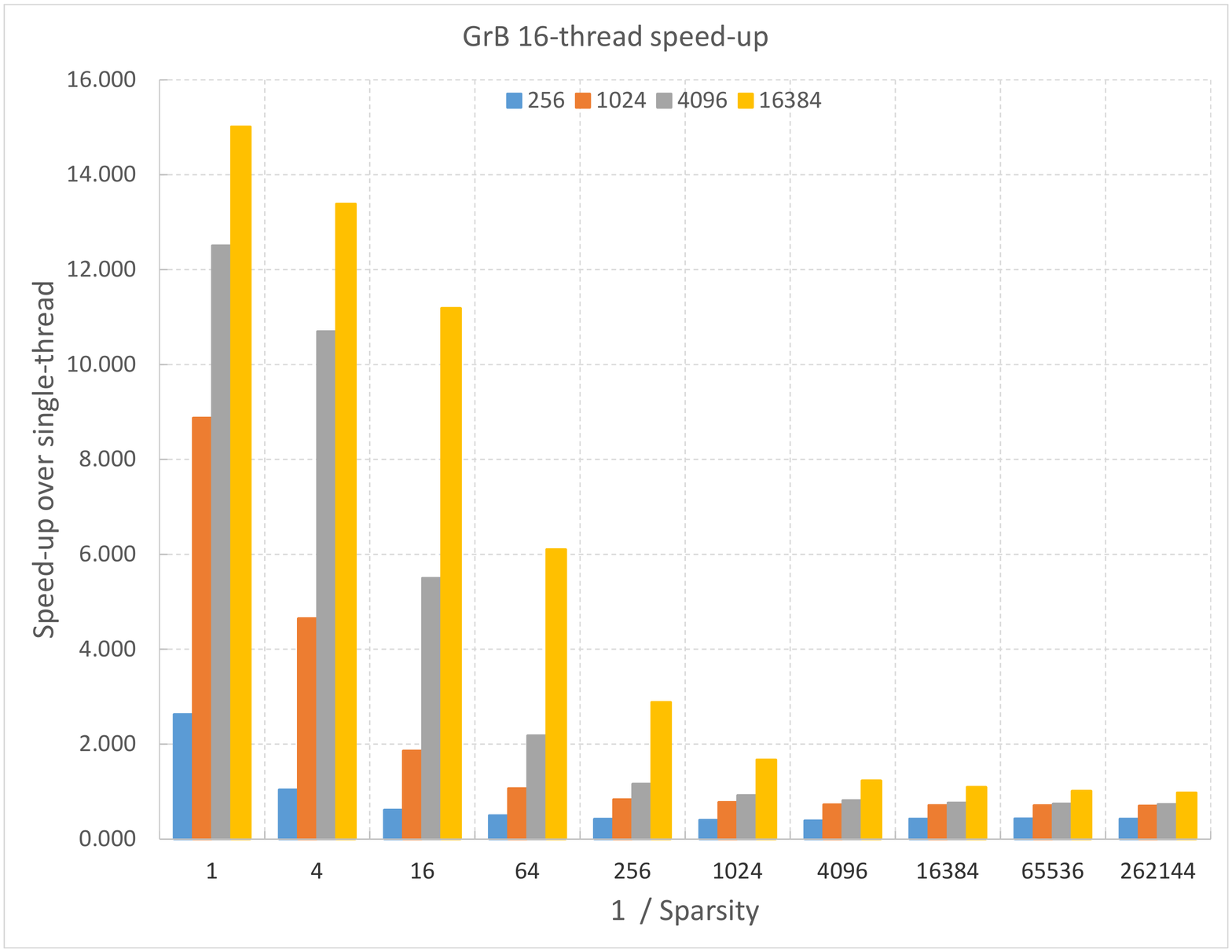} 
	\end{tabular}
	\end{center}
	\caption{Multithreaded results for GraphBLAS ({\sf GrB})
	and dense linear algebra ({\sf BLAS}) implementations of the ReLU
	DNN. Each plot is for one implementation (BLAS \emph{vs} GraphBLAS) and
	a particular level of parallelism (4 \emph{vs} 16 threads).
	Within each plot, speedup is shown as a function of inverse sparsity for 
	four different problem sizes ($256$,$1024$,$4096$,$16384$).}
	\label{Fig:OMPresults}
\end{figure*}

As expected, and consistent through the various matrix sizes, the
execution time of the dense linear algebra version is independent of the
sparsity of $\matrix{W}$.  The GraphBLAS execution time goes down with
the inverse sparsity of $\matrix{W}$.  For the measured problem sizes,
GraphBLAS is $\sim 3$ times slower than BLAS for dense matrices. The
performance advantage of BLAS starts to go down as we increase the
inverse sparsity, until the two performances approximately match for
inverse sparsity somewhat less than 4. For larger values of the inverse
sparsity, GraphBLAS shows better performance.

When matrix $\matrix{W}$ becomes very sparse (inverse sparsity $\gg n$),
the execution time of GraphBLAS levels off, as it becomes dominated
by the fixed cost of processing matrices in GraphBLAS.  That is why
the advantage of GraphBLAS over BLAS is higher for large, very sparse
matrices For $n = 32768$ and inverse sparsity $= 4096$, GraphBLAS is
$\sim 250$ times faster than the dense linear algebra implementation,
or a $1000$-fold change in relative performances from the dense case.

Equally (or maybe more) important, is the advantage in memory consumption.
A $32768 \times 32768$ matrix for dense linear algebra has approximately
1 billion elements and consumes 4~GiB of storage (for single-precision
data).  The sparse matrix in GraphBLAS has storage that is essentially
proportional to the number of nonzeros. That means that, depending on
the inverse sparsity, GraphBLAS can accommodate problem sizes that are
well beyond the means of current machines with the dense linear algebra
approach.

Figure~\ref{Fig:OMPresults} shows the speedup, as a function of inverse
 sparsity and for different sizes of the weight matrix, from
multi-threaded execution with both 4 and 16 threads,
 for the BLAS and
GraphBLAS implementations of the DNN code.
 The 4-thread measurements
for GraphBLAS were taken with the four threads bound to
 four different
cores on the same socket.  

The 16-thread measurements for GraphBLAS were taken by assigning four threads to each of the four sockets, and binding the four threads assigned to a socket to four different cores. OpenBLAS does its own thread management.

As expected, the speedup behavior for BLAS varies little with the sparsity, and whatever variation we have is more likely due to measurement noise than anything intrinsic to the parallel execution. In general, speedup is better for larger matrices, since they better amortize any fixed overhead of parallelization. 

The same behavior with respect to matrix size can be observed for the GraphBLAS implementation.  The speedup for GraphBLAS, both with 4 and 16 threads, is very good for dense matrices, but shows a noticeable drop for all matrix sizes as the inverse sparsity increases.  This is expected, since the total work that has to be executed in parallel goes down significantly as the inverse sparsity grows.

\subsection{Analysis of GraphBLAS Execution Times}

\begin{figure}
        \begin{center}
        \begin{tabular}{c}
                \includegraphics[width=3.5in]{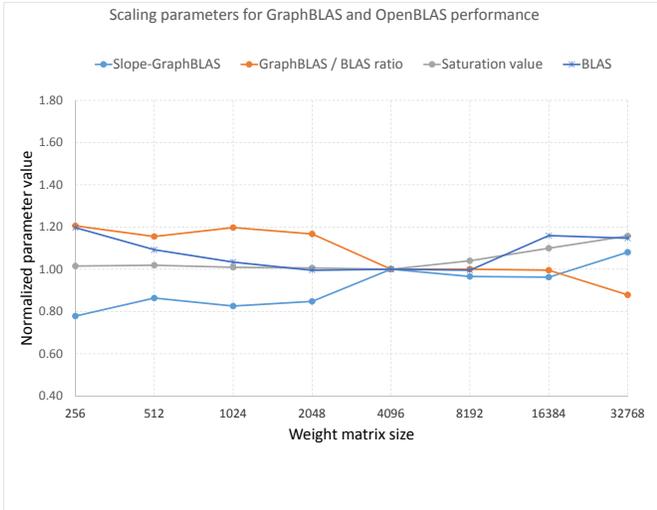}
        \end{tabular}
        \end{center}
        \caption{The various parameters of execution time curves of Figure 
	\ref{Fig:STresults}, normalized to their value at 4096*4096 matrix,
		plotted as a function of weight matrix size.
        }
        \label{Fig:STparams}
\end{figure}

The similarity in the shape of the curves for GraphBLAS performance in
Figure~\ref{Fig:STresults} suggests that some common parameters or scaling
laws define the shape of these curves.   In Figure~\ref{Fig:STparams}
we plot three such parameters and discuss their interpretation and
significance.  These are: 1) Ratio of BLAS/GraphBLAS execution times,
2) Slope of GraphBLAS execution time, and 3) Saturation value of the
execution time.  We also plot the scaling of BLAS performance normalized
to matrix size.
Since only relative performance is relevant
to reasoning about scaling, all measures are also scaled to their value
for $4096 \times 4096$ weight matrix.

The time taken by BLAS for dense weight matrices of various sizes,
normalized by the number of elements in the matrix is shown with legend
{\sf BLAS}.  As mentioned in the preceding paragraph, it is further normalized
by the time taken for a $4096 \times 4096$ matrix.  As expected, the execution
time normalized to number of elements in the matrix is almost invariant.
We believe that for matrices larger than $4096 \times 4096$, 
normalized execution time increases because data is
occasionally accessed from lower levels of cache hierarchy.  For
matrices smaller than $4096 \times 4096$,
normalized execution time is slightly higher because general overheads become
non-negligible compared to the compute cycles devoted to the kernel
calculations.

The curve with legend {\sf GraphBLAS/BLAS} ratio is the ratio of run time
for GraphBLAS implementation for $n \times n$ dense weight matrix scaled by
the BLAS run time for the same weight matrix.
This is an estimate of per element processing cost for BLAS and GraphBLAS
implementations.  For $4096 \times 4096$ matrix it is 3.2 and is relatively
invariant across weight matrix sizes.  In Figure~\ref{Fig:STresults}
it is the distance between the $y$-axis intercept of the GraphBLAS and
BLAS performance curves for a given matrix size. One can readily observe that this
ratio is inversely correlated to scaled BLAS performance.

The curve with legend {\sf Slope GraphBLAS} is an estimate of the derivative of
GraphBLAS execution time with respect to sparsity, evaluated at sparsity
value of 1.  It is computed as $(T_{S=1} - T_{S=1/4}) / 0.75 \times n^{2}$,
ratio of the difference between the execution times and the number of
non-zero elements in the weight matrix for the two sparsity values.
In this equation $S$ is the sparsity value. That is, the ratio of the number of non-zero
entries to the total number of entries.
This too is invariant across matrix sizes establishing that GraphBLAS
performance for relative dense matrices is invariant to matrix size.
This slope is the per element computing cost when the matrix has many
non-zero elements per row.  The values shown are normalized to the slope
for $4096 \times 4096$ weight matrix.

The curve with the legend {\sf Saturation value} shows the execution time of
GraphBLAS for weight matrices of sparsity $2^{-18}$, normalize by the
size (number of rows, which equals the number of columns) of the matrix.
For the $512 \times 512$ matrix this sparsity amounts to one element
in the whole matrix. For the $32768 \times 32768$ matrix, it amounts to one
element in every eight rows.  This measure approximates the overhead of
processing an almost empty matrix in GraphBLAS, and establishes it is
proportional to the number of rows/columns of the matrix.  The slightly
higher values for larger matrix sizes are probably due to the large matrices
having a larger fraction of non-empty rows.  Once again, the values shown
are normalized to the saturation value for a $4096 \times 4096$ weight matrix.

% The curve with legend Dense/Sparse $\matrix{Y}$ shows the ratio of execution time
% of dense weight matrices with dense and sparse input/output matrix $\matrix{Y}$.
% For sparse input/output matrices, the sparsity is $1/4$.  Surprisingly, the
% execution time for sparse $\matrix{Y}$ is ~33~\% higher than that for the dense
% $\matrix{Y}$ because of the extra instructions needed to handle the sparsity of $\matrix{Y}$.

\section{Conclusions}

% Matrices are a powerful tool for representing and manipulating graphs.
% Adjacency matrices represent directed-weighted-graphs with each
% row and column in the matrix representing a vertex and the values
% representing the weights of the edges.  Incidence matrices represent
% directed-weighted-multi-hyper-graphs with each row representing an edge
% and each column representing a vertex.  Perhaps the most important
% aspects of matrix-based graphs are the mathematical properties of
% commutativity, associativity, and distributivity.  These properties allow
% a very small number of matrix operations to be used to construct a large
% number of graphs.  These properties of the matrix are determined by the
% element-wise properties of addition and multiplication on the values in
% the matrix. The GraphBLAS allows these matrix properties to be readily
% applied to graphs in a low-overhead manner.

Matrix operations involving weight matrices of DNNs represent the bulk of
computation in the training and inferencing of DNNs.  As the number of
stages in the DNN and the number of nodes in each stage increase to handle
more complex classification tasks, the weight matrices will become sparse.
In this paper we have shown that the key DNN computations can be represented
in GraphBLAS, a library interface defined for sparse matrix algebra.
Furthermore, we have shown that the key step of forward propagation, with ReLU
as the nonlinearity, can be performed much more efficiently with GraphBLAS 
implementation as compared to BLAS implementation when the weight matrices 
are sparse. 

\section*{Acknowledgments}

The authors would like to thank David Bader, Ayd\i{}n Bulu{\c{c}}, Paul Burkhardt, Alan Edelman, Sterling Foster, Vijay Gadepally, John Gilbert, Timothy Mattson, Dave Martinez, Scott McMillan, Henning Meyerhenke, Victor Roytburd, and Carl Yang.
%Hedayat Alghassi, Michael Anderson, Ariful Azad, Muthu Baskaran, , Steven Dalton, Tim Davis, Joe Eaton, Alan Edelman, Sterling Foster, Vijay Gadepally, Joseph Gonzalez, Torsten Hoefler, Erik Holk, Thejaka Kanewala, Tze Meng Low,  Dave Martinez, John Matty, Asit Mishra, Samantha Misurda, Mostofa Patwary, Fabrizio Petrini, Albert Reuther, Jason Riedy, Victor Roytburd, Nadathur Satish, Narayanan Sundaram, Richard Veras, Michael Wolf, Albert-Jan Yzelman,  Peter Zhang, and Xia Zhu.



%\bibliographystyle{IEEEtran}

\bibliographystyle{IEEEtran}
\bibliography{References}

%\begin{thebibliography}{1}
%\end{thebibliography}

\end{document}